\documentclass[doublecol]{epl2}
\usepackage{color}
\pdfoutput=1
\usepackage{bm}
\usepackage{amsmath}
\usepackage{graphicx}
\usepackage{amssymb}
\usepackage{esint}

\begin{document}

\title{Enhancing the absorption of graphene in the terahertz range}

\author{N. M. R. Peres and Yu. V. Bludov }

\institute{Department of Physics and Center of Physics,
 University of  Minho, P-4710-057, Braga, Portugal, EU}


\abstract{
We study graphene on a photonic crystal operating in the
terahertz (THz) spectral range. We show that the absorption of graphene
becomes a modulated function of frequency and
can be enhanced by more than three times at specific frequency values,
depending on the parameters of the system. The problem of a semi-infinite
photonic crystal is also solved.
}

\maketitle

The optical properties of graphene have been extensively studied both
theoretically
\cite{nmrPRB06,falkovsky,stauberBZ,stauberphonons,StauberGeim,carbotte,Juan,Mishchenko,rmp,rmpPeres,LiLouie,PRL,aires,APLPhotonic}
and experimentally
\cite{nair,kuzmenko,mak,basov,Crommieopt,kuzmenko2,kuzmenkoFaraday,NatureLoh}.
More recently, the interest has shifted to how graphene interacts with electromagnetic
radiation in the THz spectral range
\cite{avouris,koppens,koppensphoto,BasovPlasmonsI,BasovPlasmonsII,Pellegrini}.
In particular, one of the goals is to enhance the absorption of graphene.
This maybe done in different ways: (i) producing micro-disks of graphene
on a layered structure \cite{avouris}; (ii) exploiting the physics of quantum
dots on graphene \cite{koppensphoto};
(iii) using a graphene based
grating \cite{nunoSPP,YuliyPRB};
 (iv) putting graphene inside an
optical cavity \cite{PeresCavity,MuellerCavity};  and (v)
depositing graphene on a photonic crystal, operating in the visible range of the
spectrum \cite{APLPhotonic}.
In cases (i), (ii), and (iii) the excitation of plasmons is responsible for the enhancement of the absorption. In case (iv)
the radiation performs many round trips inside the cavity enhancing the chances
of being absorbed by graphene. In case (v) the authors use a photonic crystal
made of SiO$_2$/Si. In the visible range of the spectrum the dielectric constants
of SiO$_2$ and Si differ by more than one order of magnitude and choosing the
width of the  SiO$_2$/Si appropriately it is possible to induce a large photonic
band gap in the visible range. Combining the presence of the band gap
with an initial spacer layer the absorption can be enhanced by a factor of four.
In the case studied in Ref. \cite{APLPhotonic} the optical conductivity of
graphene is controlled by inter-band transitions. In this regime,
the conductivity is frequency
independent and reads\cite{nair}
\begin{equation}
\sigma_0=\frac{\pi e^2}{2h}\,,
\end{equation}
a value dubbed the universal conductivity of graphene.

In general, the conductivity of graphene is a sum of two contributions: (i) a Drude term,
describing intra-band processes and (ii) a term describing inter-band transitions. At zero
temperature the optical conductivity has a simple analytical expression
\cite{nmrPRB06,falkovsky,rmp,rmpPeres,StauberGeim}. For what concerns our study,
the physics of the system is dominated by the Drude contribution,
which reads
\begin{equation}
\sigma_D =\sigma_0\frac{4\epsilon_F}{\pi}\frac{1}{\hbar\Gamma-i\hbar\omega}\,,
\label{eq_sigma_xx_semiclass}
\end{equation}
 where
$\Gamma$ is the relaxation rate, $\epsilon_F>0$ is the
Fermi level position with respect to the Dirac point, and $\omega$ is the
frequency of the incoming radiation. It should be noted that $\sigma_D$
has a strong frequency dependence and is responsible for the optical behaviour
of graphene in the THz spectral range.

\begin{figure}[htb]
 \begin{center}
 \includegraphics*[width=8cm]{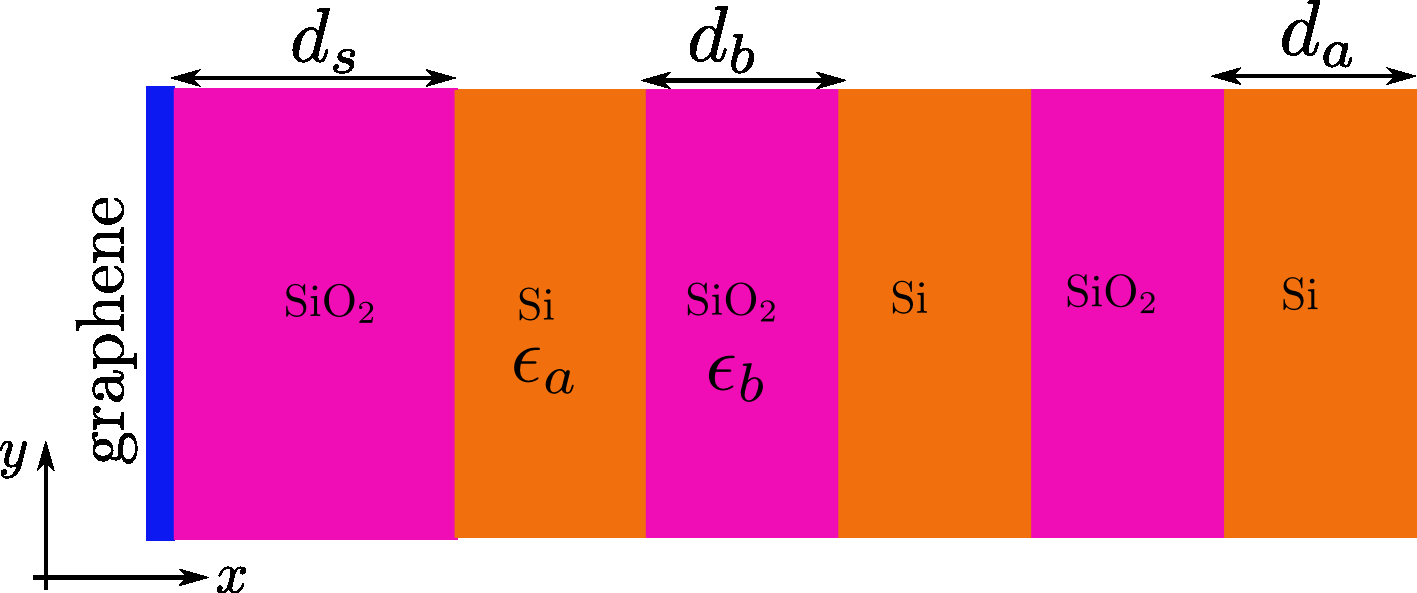}
 \end{center}
\caption{Graphene on a micro-structure made of SiO$_2$/Si. The initial spacer
 has a different length, $d_s$, from the other
layers of SiO$_2$ and Si, which have length $d_j$.  Graphene can be doped either using a transparent  top gate  or chemically
\cite{chemicaldopeII,chemicaldopeI}; in the latter case no gate is necessary.}
\label{fig_device}
\end{figure}

In what follows, the absorption of graphene on a system such as the one
represented in Fig. \ref{fig_device} is computed. The micro-structure
has a first layer (spacer) of SiO$_2$, of length $d_s$, followed by
 a periodic array of Si/SiO$_2$ which extends for several layers.
We have found that no more than few layers are necessary for obtaining an enhancement of the
absorption. We write $d_s$ as
\begin{equation}
 d_s = \frac{\lambda}{\beta\sqrt{\epsilon_s}}\,,
\end{equation}
where $\beta$ is a numerical parameter chosen to optimize the absorption,
$\epsilon_s$ is the dielectric constant of the spacer (SiO$_2$),
and $\lambda$ is a parameter of the order of few microns also
chosen to  optimize the absorption. The lengths  $d_j$ of the Si/SiO$_2$
periodic structure layers are given by
\begin{equation}
d_j=  \frac{\lambda}{\sqrt{\epsilon_j}}\,,
\end{equation}
where $j=a,b$, with $\epsilon_a$($\epsilon_b$) standing for the dielectric constant of
Si(SiO$_2$). In the THz we approximate the dielectric constants of Si and  SiO$_2$
by their static  values, $\epsilon_a=11.9$ and $\epsilon_b=3.9$.

To determine the absorption we have to compute the amount of light reflected
and transmitted by and through the structure, respectively. If we denote the
transmittance by $\cal T$ and the reflectance by $\cal R$, the absorbance
is defined, assuming normal incidence, as
\begin{equation}
 {\cal A}=1- {\cal T}- {\cal R}\,.
\end{equation}
It is also assumed that the radiation impinges from the air and emerges the structure
to the air as well. Since we are considering normal incidence, no distinction is necessary
between $s-$ and $p-$polarized waves.

An electromagnetic wave impinges on graphene from the left along the
$x-$direction. Due to reflection on the structure the electric field has the form
\begin{equation}
 \bm E = (E_{0+}e^{ikx} +E_{0-}e^{-ikx})\textbf{u}_y\,.
\end{equation}
From Maxwell's equations, the magnetic field is given by
\begin{equation}
 i\omega B_z = \partial_x E_y\,.
\end{equation}
At the graphene interface the boundary conditions are
$E_L=E_R$ and $B_L-B_R=\mu_0\sigma_DE_L$, where the subscript $L$($R$)
stands for the fields to the left(right) of graphene. At the other interfaces
the boundary condition for the magnetic field is simply
$B_L=B_R$. On the right hand side of the structure we write the electric
field as
\begin{equation}
 \bm E = E_{t+}e^{ikx}\textbf{u}_y\,.
\end{equation}
Using the transfer matrix method we can relate the fields on the left of the
structure to the fields on the right as
\begin{equation}
 \left[
\begin{array}{c}
 E_{0+}\\
E_{0-}
\end{array}
\right]=\hat{\mathbb{T}}
 \left[
\begin{array}{c}
 E_{t+}\\
 0
\end{array}
\right]=
 \left[
\begin{array}{cc}
 t_{11} & t_{12}\\
t_{21} & t_{22}
\end{array}
\right]
 \left[
\begin{array}{c}
 E_{t+}\\
 0
\end{array}
\right]\,,
\end{equation}
from where it follows
\begin{equation}
 {\cal T}=\frac{\left|E_{t+}\right|^2}{\left|E_{0+}\right|^2}=\frac{1}{\vert t_{11}\vert^2}\,,
\end{equation}
and
\begin{equation}
 {\cal R}=\frac{\left|E_{0-}\right|^2}{\left|E_{0+}\right|^2}=\frac{\vert t_{21}\vert^2}{\vert t_{11}\vert^2}\,.
\end{equation}
For the structure we are considering, the transfer matrix $\mathbb{T}$
is given by
\begin{equation}
 \hat{\mathbb{T}} = (\hat{K}_{0,b}+\alpha f \hat{J})\hat{\Phi}_s\hat{K}_{b,a}\hat{T}^N\hat{\Phi}_a\hat{K}_{a,0}\,,
\end{equation}
where $N$ is the number of times the matrix $\hat{T}$ is repeated in the structure
and $\hat{T}$ reads
\begin{equation}
 \hat{T}=\hat{\Phi}_a\hat{K}_{a,b}\hat{\Phi}_b\hat{K}_{b,a}\,.
\end{equation}
The several matrices appearing in $\mathbb{T}$ and $\hat{T}$ are given below.
The matrix characterizing the free propagation through a given material
is
\begin{equation}
 \hat{\Phi}_j = \left[
\begin{array}{cc}
 e^{-i\omega\sqrt{\epsilon_j}d_j/c}& 0\\
0 & e^{i\omega\sqrt{\epsilon_j}d_j/c}
\end{array}
\right]\,,
\end{equation}
with $j=s,a,b$. The matrix referring to the transfer through the boundary of two dielectrics
is given by
\begin{equation}
 \hat{K}_{i,j}=
\left[
\begin{array}{cc}
 (\sqrt{\epsilon_j/\epsilon_i}+1)/2& (1-\sqrt{\epsilon_j/\epsilon_i})/2\\
(1-\sqrt{\epsilon_j/\epsilon_i})/2&(\sqrt{\epsilon_j/\epsilon_i}+1)/2
\label{eq_prodt_Ks}
\end{array}
\right]\,.
\end{equation}
In particular, $K_{a,0}$ and $K_{0,b}$ refers to the passage from the air to the last slab of Si and from the first SiO$_2$ spacer
to the air, respectively [thus $\epsilon_0=1$ in Eq. (\ref{eq_prodt_Ks})].
The matrix $K_{0,b}+\alpha f J$ characterizes the transfer
through graphene to air. Here matrix
\begin{equation}
 \hat{J} = \left[
\begin{array}{cc}
 1& 1\\
-1&-1
\end{array}
\right]\,,
\end{equation}
$\alpha\approx1/137$ is the fine structure constant and $f$ is a dimensionless function defined
by
\begin{equation}
 f=\frac{2\epsilon_F}{\hbar\Gamma -i\hbar\omega}\,.
\end{equation}
We note that in the infinite structure we can apply Bloch theorem, and relate the electric field amplitudes in the unit cell $m$ with those in the
unit cell $m+1$ using the transfer matrix $T$ as
\begin{equation}
 \left[
\begin{array}{c}
 E_{a+}^m\\
 E_{a-}^m
\end{array}
\right]=
\hat{T}\left[
\begin{array}{c}
 E_{a+}^{m+1}\\
 E_{a-}^{m+1}
\end{array}
\right]=\hat{T}e^{iq(d_a+d_b)}
\left[
\begin{array}{c}
 E_{a+}^m\\
 E_{a-}^m
\end{array}
\right]\,,
\label{eq:bloch}
\end{equation}
where $q$ is the Bloch wave vector. Then the spectrum of the infinite structure can be obtained from the condition ${\rm det}|\hat{I}-\hat{T}e^{iq(d_a+d_b)}|=0$ (here $\hat{I}$ is the unit matrix). Using the property of transfer matrix ${\rm det}|\hat{T}|=1$, we obtain that
the spectrum of the infinite photonic crystal,
\begin{equation}
 {\rm Tr}(\hat{T})=2\cos[(d_a+d_b)q]\,,
\label{eq_trace}
\end{equation}
is defined by trace of $\hat{T}$ only.
From Eq. (\ref{eq_trace})  the photonic band gaps can be obtained
with no difficulty. We can also access the photonic band gaps by transmittance calculations,
a route we choose below.

\begin{figure}[htb]
 \begin{center}
 \includegraphics*[width=6.5cm]{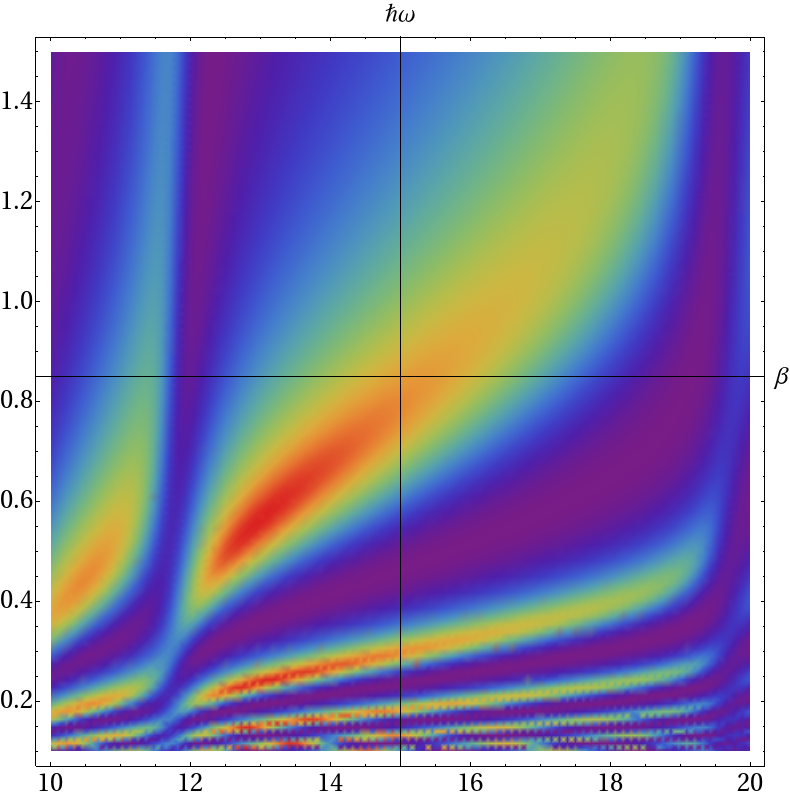}

\includegraphics*[width=6.5cm]{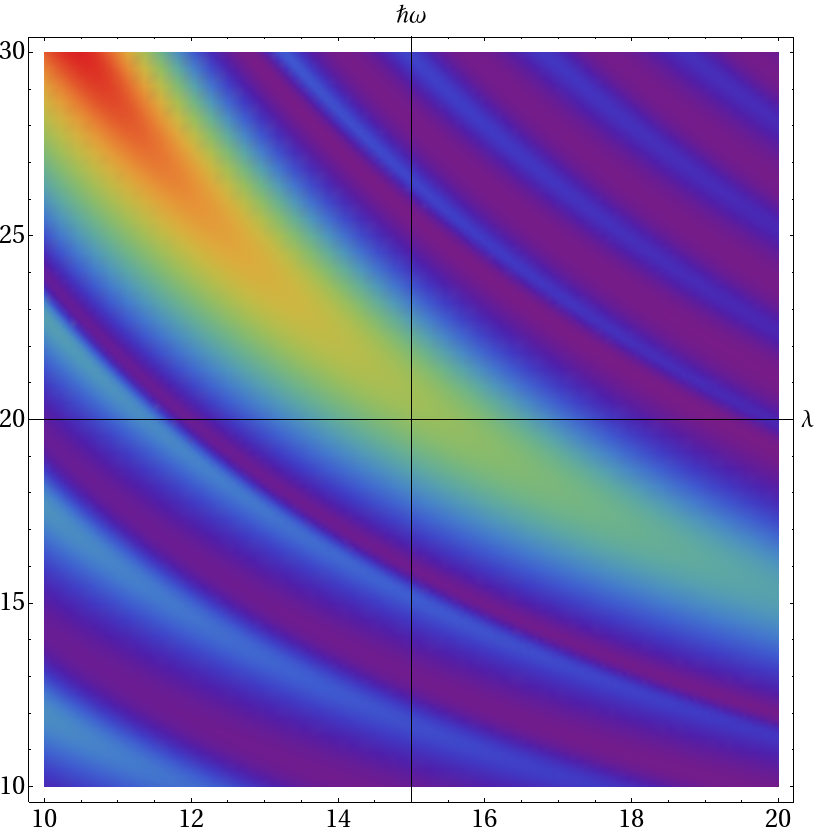}
 \end{center}
\caption{Intensity plot of the absorbance. Top panel: as function of the parameter
$\beta$ and the photon energy $\hbar\omega$ (in meV). Bottom panel:
as function of $\lambda$ (in $\mu$m) and $\hbar\omega$. High intensity spots
corresponds points in the parameter space where the absorbance attains
its higher values. We have used $N=4$, $\beta=0.85$, $\lambda=$20 $\mu$m,
$\hbar\Gamma=2.5$ meV, and  $\epsilon_F=0.25$ eV,
a typical value. For reference, 4.1 meV corresponds to 1 THz.}
\label{fig_intensity_plot}
\end{figure}
\begin{figure}[!htb]
 \begin{center}
 \includegraphics*[width=7cm]{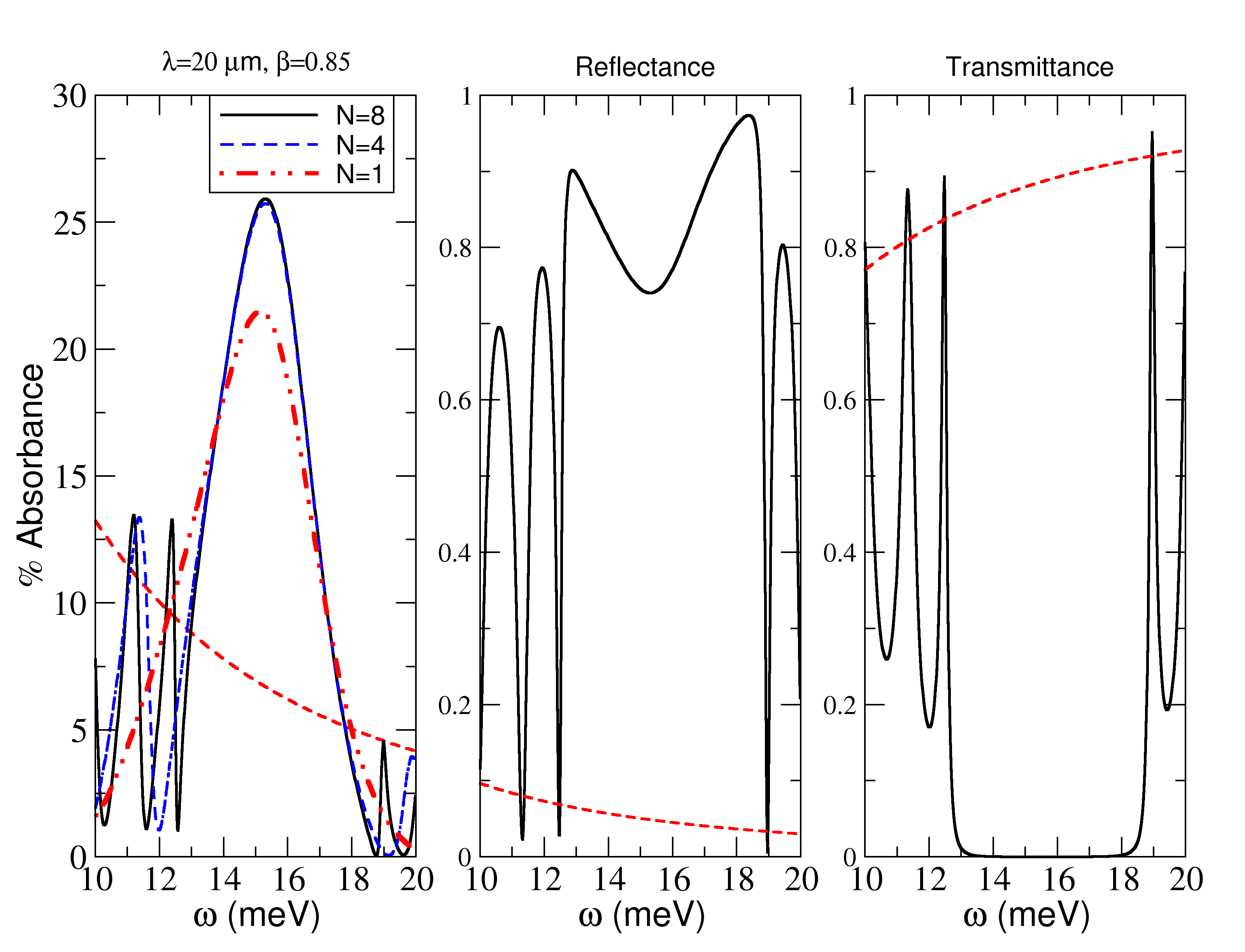}
 \end{center}
\caption{Absorbance (left), reflectance (center), and
transmittance (right) as function of frequency. The dashed line
represents the same quantity but for free-standing graphene.
 We have used $N=8$, $\beta=0.85$, $\lambda=$20 $\mu$m,
$\hbar\Gamma=2.5$ meV, and  $\epsilon_F=0.25$ eV.
The cases $N=1,4$ are also shown in the absorbance panel.}
\label{fig_THZ}
\end{figure}

In Fig. \ref{fig_intensity_plot} we represent the absorbance as function of: (i) the parameter
$\beta$ and the photon energy $\hbar\omega$ (in meV)[top panel];
(ii) the parameter $\lambda$ (in $\mu$m) and $\hbar\omega$ [bottom panel].
Recall that the
parameter $\beta$ controls the  length of the SiO$_2$ spacer whereas $\lambda$
influences the length of the layers constituting the periodic structure.
The high-intensity spots correspond to regions in parameters space where
the absorbance is higher. The axes in Fig. \ref{fig_intensity_plot} intercept
at one of those high intensity spots. The width of the high intensity regions along the
$\hbar\omega$ direction gives the width of the resonance. The position of the
high-intensity regions is   sensitive to the parameter $\lambda$, as can be seen in the bottom panel
of Fig. \ref{fig_intensity_plot}. These type of plots allow the optimization of the structure in order to
 maximize the absorbance. An optimized absorbance peak is given in  Fig. \ref{fig_THZ}.
The parameters chosen are those corresponding to the interception of the horizontal and vertical lines in Fig. \ref{fig_intensity_plot}.
In Fig. \ref{fig_THZ} we represent the absorbance, the reflectance, and
the transmittance as function of the frequency of the incoming photon.
Since the dielectric constants we have used for SiO$_2$ and Si
are real, all the absorption comes of graphene alone. The intricate
behaviour
seen on the reflectance and transmittance curves, on the other hand,
is mainly due to the structure on the right hand side of graphene. The dashed lines
in the same figure represent the behaviour of free standing graphene.

Several aspects are worth mention. Although we have used a fairly small
value of $N$, a stop gap due to the photonic band gap of the infinite
crystal is already apparent. This happens because the difference between the
dielectric constants of SiO$_2$ and Si is significant.  Most important however
is the enhancement of the absorbance near the center of the photonic
band gap. We see that around an energy $\hbar\omega\approx15.3$ meV
the absorbance is almost four times larger than the absorbance of free standing graphene
(assuming it can be doped some how, e. g., by chemical doping \cite{chemicaldopeII,chemicaldopeI};
if we had chosen graphene on a SiO$_2$ wafer of the size of the
photonic crystal the absorbance would be smaller than that for the free standing case).
If we had chosen somewhat different
parameters the resonance would shift to the edges of the stop gap.
In the absorbance panel we plot the cases  $N=1$, $N=4$, and $N=8$.
As it can be seen, for $N=1$ the absorbance is not as intense as in the other two cases. At the same time the main
difference is the appearance of new satellite  peaks to the left and to the right of the main
one when we change from $N=1$ to $N=4$, and to $N=8$. The nature of these peaks can be expressed in the following simplified manner.
In the infinite periodic structure allowed bands are  continuous functions of Bloch wave vector $q$, while in the periodic structure containing a finite number of periods $N$ the effect of quantization takes place: allowed bands consist of the discrete set of the frequencies, which can be determined from Eq.(\ref{eq_trace}) using the discrete Bloch vectors $q_n=n\pi/(N+1)$ for $n\in[1,N]$ (see Ref. \cite{Bludov2007}). In particular, for $N=4$ the mode with $n=4$ correspond to frequencies $\omega\approx 11.4\,$meV and $\omega\approx 19.6\,$meV (here we take into account the frequency range of Fig.\ref{fig_THZ} only), while for $N=8$ the allowed frequencies are
$11.2\,$meV (mode with $n=7$),
$12.3\,$meV, and
$18.8\,$meV (mode with $n=8$).
In both cases the predicted frequencies  correspond well with the absorption peaks below and above the stop gap
seen in Fig.\ref{fig_THZ}.

\begin{figure}[t]
 \begin{center}
 \includegraphics*[width=7cm]{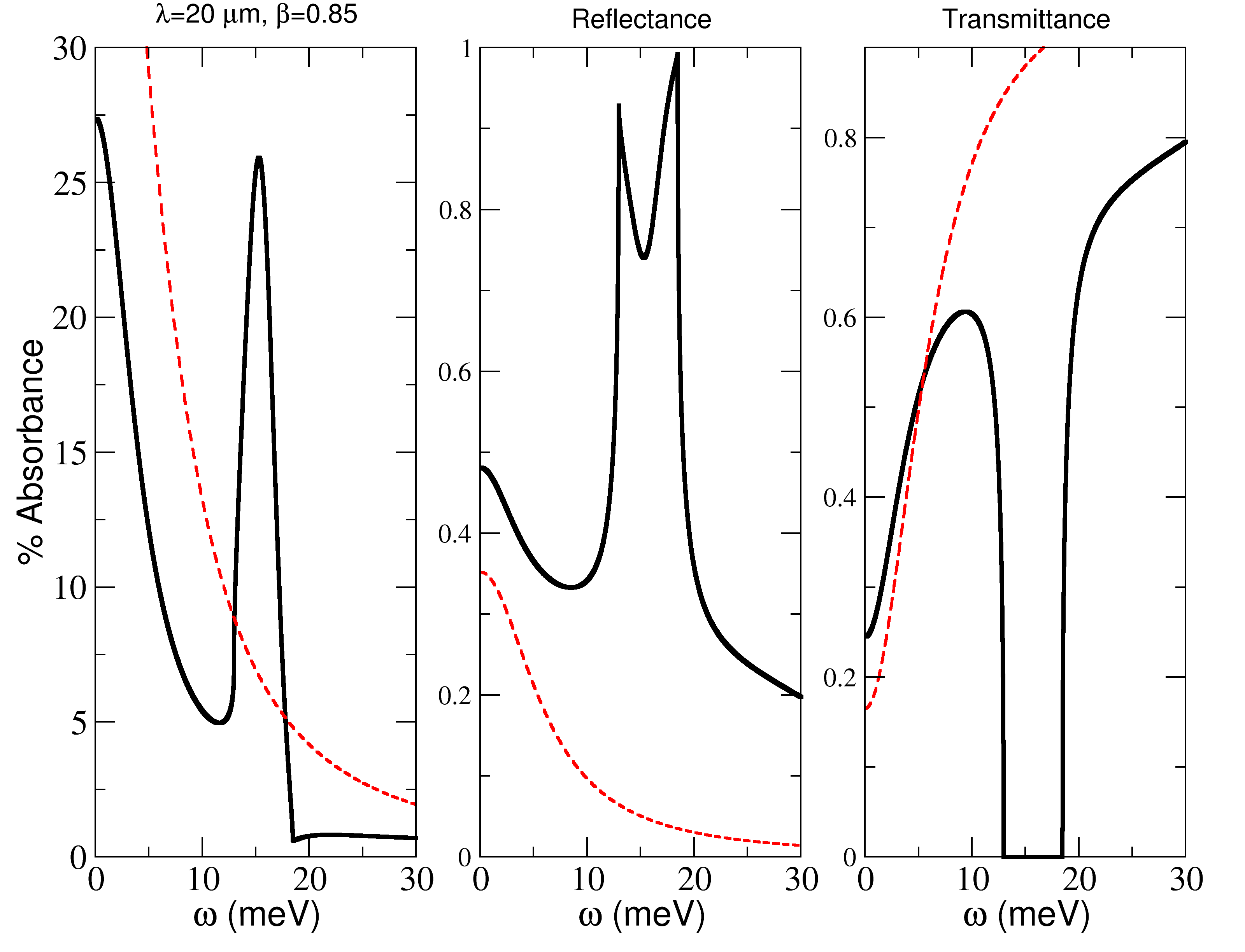}
 \end{center}
\caption{Absorbance (left), reflectance (center), and
transmittance (right) as function of frequency for a semi-infinite
photonic crystal. The dashed line
represents the same quantity but for free-standing graphene.
 We have used  $\beta=0.85$, $\lambda=$20 $\mu$m,
$\hbar\Gamma=2.5$ meV, and  $\epsilon_F=0.25$ eV.}
\label{fig_THZ_semi_infinite}
\end{figure}

It is instructive to compute the absorbance, reflectance, and transmittance
when the photonic crystal is semi-infinite. In this case the oscillations
due to the finiteness of the crystal seen in Fig. \ref{fig_THZ} are absent
and it is possible to derive  simple analytical formulae for the above quantities.
From Eq.(\ref{eq:bloch})
it follows that
\begin{equation}
 E^m_{a-}=-\frac{T_{11}-e^{-iq(d_a+d_b)}}{T_{12}}E^m_{a+}\equiv\rho(q)E^m_{a+}\,,
\end{equation}
where $T_{ij}$ are the matrix elements of the matrix $\hat{T}$.
Now the amplitudes for $x<0$ relate to the amplitudes in the first
Si slab as
\begin{eqnarray}
 \left[
\begin{array}{c}
 E_{0+}\\
 E_{0-}
\end{array}
\right]&=&
(\hat{K}_{0,b}+\alpha f\hat{J})\hat{\Phi}_s\hat{K}_{b,a}
\left[
\begin{array}{c}
 E_{a+}^1\\
 E_{a-}^1
\end{array}
\right]\nonumber\\
&=&
\left[
\begin{array}{cc}
 M_{11} & M_{12}\\
 M_{21} & M_{22}
\end{array}
\right]
\left[
\begin{array}{c}
 E_{a+}^1\\
 E_{a-}^1
\end{array}
\right]
\,,
\end{eqnarray}
from which it follows that
\begin{eqnarray}
 E_{0+}=(M_{11}+\rho(q)M_{12})E^1_{a+}\,,\\
 E_{0-}=(M_{21}+\rho(q)M_{22})E^1_{a+}\,,
\end{eqnarray}
leading to the reflectance amplitude
\begin{equation}
 r=\frac{E_{0-}}{E_{0+}}=\frac{M_{21}+\rho(q)M_{22}}{M_{11}+\rho(q)M_{12}}\,,
\end{equation}
from which the reflectance is computed as ${\cal R}=\vert r\vert^2$.
Because in each slab there is always an incoming and an outgoing wave the
calculation of the transmittance is less obvious. We have to compute
two field ratios:
\begin{equation}
 t_+=\frac{E^1_{a+}}{E_{0+}}=\frac{1}{M_{11}+\rho(q)M_{12}}\,,
\end{equation}
and
\begin{equation}
 t_-=\frac{E^1_{a-}}{E_{0+}}=\frac{\rho(q)}{M_{11}+\rho(q)M_{12}}\,.
\end{equation}
Then, the transmittance is given by
\begin{equation}
 {\cal T} = \sqrt{\epsilon_a}(\vert t_+\vert^2-\vert t_-\vert^2)\,,
\end{equation}
where the factor $\sqrt{\epsilon_a}$ comes from the calculation of
the flux. In Fig. \ref{fig_THZ_semi_infinite} we plot the
absorbance (left), reflectance (center), and
transmittance (right) as function of frequency for a semi-infinite
photonic crystal. The resonance in the absorbance coincides exactly with that seen in
Fig. \ref{fig_THZ}, but the oscillations also seen in Fig. \ref{fig_THZ}
are not present here.

In conclusion, it is possible to use a photonic crystal-like structure
to enhance the absorption of radiation by graphene in the THz spectral range at
specific frequencies.
The effect depends on the geometric properties of the crystal.
At the center
of the photonic band gap the absorption increases at the expenses of the
reflectance (the resonance can shift its position depending on the parameters
chosen). If the Fermi energy increases so does the absorption and the
width of the peak. It is also clear that for a finite crystal the absorbance becomes a modulated
function of  frequency. We have chosen the widths of the SiO$_2$ and Si
having the same dependence on $\lambda$ (exception made to the initial layer
which has the extra parameter $\beta$).
If we had chosen them to be different we would have an extra degree of freedom
to tweak the absorbance curves, but the optimization procedure would become
cumbersome. If we had sandwiched graphene between two dielectrics the absorbance
would decrease because of the optical conductivity would appear in the equations
divided by the square root of the dielectric constant of the medium. For the parameters
of Fig. \ref{fig_THZ} the next stop gap appears at an energy of about 47 meV (not shown).
There  the enhancement of the absorbance is larger by a factor of four relatively
to the free standing case.  Finally we note that the device does not necessarily require a gate for
doping graphene. Stable hole doping by chemical methods have recently been achieved, without
affecting the transparency of the material \cite{chemicaldopeII}.

\providecommand{\newblock}{}

\end{document}